\documentclass[aps,prl,twocolumn,showpacs,amsmath,amssymb,superscriptaddress]{revtex4}
\usepackage{graphicx}
\usepackage{amssymb}
\usepackage{natbib}

\begin{document}

\title{Electronic nematic correlations in the stress free tetragonal state of BaFe$_{2-x}$Ni$_{x}$As$_{2}$}
\author{Haoran Man}
\affiliation{Department of Physics and Astronomy, Rice University, Houston, Texas 77005, USA}
\author{Xingye Lu}
\affiliation{Department of Physics and Astronomy, Rice University, Houston, Texas 77005, USA}
\affiliation{Beijing National Laboratory for Condensed Matter
Physics, Institute of Physics, Chinese Academy of Sciences, Beijing
100190, China}
\author{Justin S. Chen}
\affiliation{Department of Physics and Astronomy, Rice University, Houston, Texas 77005, USA}
\author{Rui Zhang}
\affiliation{Department of Physics and Astronomy, Rice University, Houston, Texas 77005, USA}
\author{Wenliang Zhang}
\affiliation{Beijing National Laboratory for Condensed Matter
Physics, Institute of Physics, Chinese Academy of Sciences, Beijing
100190, China}
\author{Huiqian Luo}
\affiliation{Beijing National Laboratory for Condensed Matter
Physics, Institute of Physics, Chinese Academy of Sciences, Beijing
100190, China}
\author{J. Kulda}
\affiliation{Institut Laue-Langevin, 6, rue Jules Horowitz, BP 156, 38042 Grenoble Cedex 9, France}
\author{A. Ivanov}
\affiliation{Institut Laue-Langevin, 6, rue Jules Horowitz, BP 156, 38042 Grenoble Cedex 9, France}
\author{T. Keller}
\affiliation{Max-Planck-Institut f$\ddot{u}$r Festk$\ddot{o}$rperforschung, Heisenbergstrasse 1, D-70569 Stuttgart, Germany}
\affiliation{Max Planck Society Outstation at the Forschungsneutronenquelle Heinz Maier-Leibnitz (MLZ), D-85747 Garching, Germany}
\author{Emilia Morosan}
\affiliation{Department of Physics and Astronomy, Rice University, Houston, Texas 77005, USA}
\author{Qimiao Si}
\affiliation{Department of Physics and Astronomy, Rice University, Houston, Texas 77005, USA}
\author{Pengcheng Dai}
\email{pdai@rice.edu} 
\affiliation{Department of Physics and Astronomy, Rice University, Houston, Texas 77005, USA}

\begin{abstract}
{We use transport and neutron scattering to study electronic, structural, and magnetic properties of 
the electron-doped BaFe$_{2-x}$Ni$_x$As$_2$ iron pnictides in the external stress free detwinned state.
Using a specially designed in-situ
mechanical detwinning device, we demonstrate that the in-plane resistivity
anisotropy observed in the uniaxial strained tetragonal state of BaFe$_{2-x}$Ni$_x$As$_2$ 
below a temperature $T^\ast$, previously identified as a signature of the electronic nematic phase, 
is also present in the stress free tetragonal phase below $T^{\ast\ast}$ ($<T^\ast$).  
By carrying out neutron scattering measurements on BaFe$_2$As$_2$ and BaFe$_{1.97}$Ni$_{0.03}$As$_2$, 
we argue that the resistivity anisotropy in 
the stress free tetragonal state of iron pnictides arises from the magnetoelastic coupling associated with 
antiferromagnetic order.  These results thus 
indicate that the local lattice distortion and nematic spin correlations are responsible for the resistivity anisotropy in the tetragonal
state of iron pnictides.}
\end{abstract}


\maketitle

There is growing experimental evidence suggesting that the electronic nematic phase,
a translationally invariant metallic phase (satisfy the $90^\circ$-rotational or $C_4$ symmetry) 
with a spontaneously generated spatial electronic
anisotropy, is intimately connected with high-transition (high-$T_c$) temperature superconductivity \cite{fradkin}. 
For iron pnictide superconductors such as BaFe$_{2-x}T_x$As$_2$ ($T=$ Co, Ni) \cite{kamihara,stewart,dai,cruz,qhunag,mgkim}, 
their parent compound BaFe$_2$As$_2$ exhibits a tetragonal to orthorhombic structural phase transition at temperature $T_s$, followed by a
paramagnetic to antiferromagnetic (AF) phase transition at 
$T_N$ ($T_s\ge T_N$) with a collinear AF structure [Fig. 1(a)] \cite{cruz,qhunag,mgkim}.
Upon electron-doping via Co or Ni substitution for Fe, the $T_N$ and $T_s$ are gradually Figure1
Figure2
Figure3
Figure4
Supp1
Supp2
Supp3ressed 
and optimal superconductivity emerges 
near $x\approx 0.1$ for BaFe$_{2-x}$Ni$_x$As$_2$ \cite{CLester09,Nandi,Yoshizawa12,HQLuo12,XYLu13a}.
Due to the formation of twin domains in the orthorhombic state of  BaFe$_{2-x}T_x$As$_2$ below $T_s$, 
the intrinsic electronic properties of these materials can be probed by applying a uniaxial 
pressure (strain) along one-axis of the orthorhombic lattice to detwin the single crystal \cite{jhchu,matanatar,fisher,chu12}.
While there is indeed a large in-plane resistivity anisotropy in the uniaxial strain detwinned 
BaFe$_{2-x}T_x$As$_2$ below $T_s$, the anisotropy persists in the paramagnetic tetragonal state 
below a characteristic temperature $T^\ast$ ($T^\ast >T_s\ge T_N$), thus suggesting the presence of 
electronic nematic correlations above $T_s$ and below $T^\ast$ \cite{jhchu,matanatar,fisher,chu12,HHKuo,HHKuo15}. 
However, since the uniaxial strain necessary to detwin the sample also 
enhances $T_N$ \cite{Dhital} and introduces an explicit symmetry breaking field, it is unclear 
if there will be resistivity anisotropy in the stress free tetragonal state below $T^\ast$
upon releasing the applied external uniaxial strain.
From transport \cite{chu12,HHKuo,HHKuo15}, inelastic neutron scattering \cite{xylu14}, 
and thermodynamic measurements \cite{XLuo15}, $T^\ast$ is believed to mark 
a temperature range of nematic fluctuations with structure and magnetic phase transitions occurring at $T_s$ and $T_N$, respectively.
On the other hand, magnetic torque and X-ray diffraction experiments on stress free samples of BaFe$_2$As$_2$ suggest 
that $T^\ast$ is a signature of a ``true'' second-order nametic phase transition
from the high-temperature tetragonal phase to a low-energy orthorhombic phase \cite{Kasahara12}. 
To understand the role of electronic nematic phase in high-$T_c$ superconductivity, it is important to reveal the origin of 
the resistivity anisotropy above $T_s$ without external uniaxial strain and determine 
the nature of the nematic correlations below $T^\ast$ \cite{fernandes11}.

\begin{figure}[t]
\includegraphics[scale=.5]{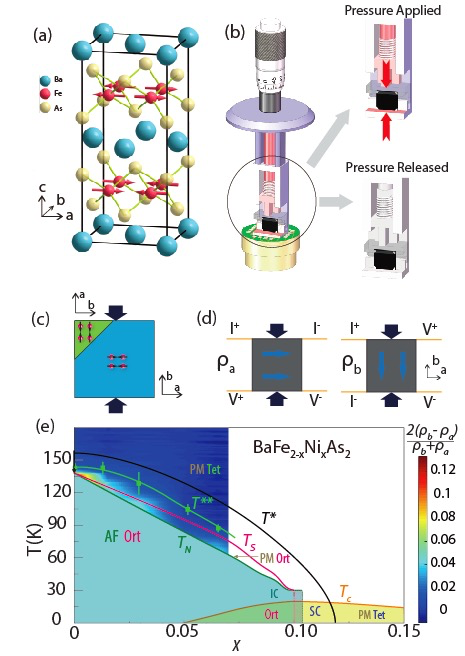}
\caption{Summary of transport and neutron scattering results.
 (a) The crystal and magnetic structures of BaFe$_2$As$_2$ 
in the AF orthorhombic state where the arrows mark the moment directions of iron \cite{qhunag}.
 (b) Schematic diagram the device used to change pressure on the sample in-situ. A micrometer is used to adjust the length of the spring and therefore the pressure applied on the sample. 
The applied pressure then can be released by fully retreat of the micrometer, as indicated in the expanded schematic on the right.
(c) The uniaxial strain is applied along the $b$-axis of the crystal, enlarging the blue domain and reducing the green domain. 
(d) Wire connection and current flow directions for resistivity measurements using Montgomery method. The black arrows 
indicate the uniaxial pressure direction and the blue arrows in the sample are the current direction for each set up.
(e) The electronic phase diagram of BaFe$_{2-x}$Ni$_x$As$_2$ as a function of Ni-doping as determined from 
previous experiments \cite{XYLu13a}. The spin excitation anisotropy temperatures under uniaxial strain are marked as $T^\ast$ \cite{xylu14}.
The AF orthorhombic (Ort), incommensurate AF (IC) \cite{XYLu13a}, paramagnetic tetragonal (PM Tet),
superconductivity (SC) phases are clearly marked. 
$T^{\ast\ast}$ marks the temperature below which resistivity anisotropy appears in the strain free tetragonal state.
}
\end{figure}

We use transport and neutron scattering to study the resistivity anisotropy, magnetic order, 
and lattice distortion in parent compound 
BaFe$_2$As$_2$ ($T_N\approx T_s \approx 138$ K) and electron-doped BaFe$_{2-x}$Ni$_x$As$_2$ 
($x=0.015, 0.03, 0.05, 0.065$). In previous transport and neutron scattering measurements, 
the applied uniaxial pressure necessary to detwin the crystal 
in the orthorhombic AF phase remains in the paramagnetic tetragonal state  ($T>T_N, T_s$), 
thus complicating the interpretation of the observed in-plane resistivity
and spin excitation anisotropy \cite{jhchu,matanatar,chu12,HHKuo,HHKuo15,Dhital,xylu14}. 
To avoid this problem, we have designed an in-situ mechanical sample clamp which can apply and release uniaxial
pressure at any temperature, similar to the device used to study the anisotropic optical 
response in iron pnictides \cite{Mirri}. Figure 1(b) shows the schematics of the
sample stick with a micrometer on the top. The magnitude of the uniaxial pressure along the
$b$-axis direction of the orthorhombic lattice 
is applied by a spring that is controlled by the 
displacement of the micrometer (and external applied pressure) [Fig. 1(c)]. By applying uniaxial pressure at room 
temperature (above $T_N$ and $T_s$), cooling the sample to below $T_N$, and then releasing the pressure, we can 
in principle obtain the 
single domain sample without external strain (stress free).  
To conclusively determine the sample detwinning ratio and compare them with the resistivity anisotropy measurement, we 
used two original sample sticks one for transport in a physical property measurement system (PPMS) and one for neutron scattering experiment on IN8 triple-axis spectrometer at Institut 
Laue-Langevin (ILL). Our key finding is that the resistivity anisotropy 
in BaFe$_{2-x}$Ni$_x$As$_2$ seen in the uniaxial strained tetragonal phase 
below $T^\ast$ is also present in the stress free tetragonal state, but at a lower temperature $T^{\ast\ast}<T^\ast$ [Fig. 1(e)].
In addition, our neutron Larmor diffraction measurements \cite{Rekveldt2001,Pfleiderer} on temperature dependence of the lattice spacing ($d$) and its distortion ($\Delta d$) in lightly 
electron-doped iron pnictides reveal that the lattice distortion  
increases on cooling, passes smoothly across $T_s$, and enhances dramatically on approaching $T_N$ with no 
observable anomaly above $T_s$.  
These results suggest that the resistivity anisotropy observed in the external uniaxial pressure free tetragonal state 
of BaFe$_{2-x}$Ni$_x$As$_2$ arises from a strong magnetoelastic coupling induced by AF order, and there are 
no additional thermodynamic phase transitions above $T_s$ \cite{Fernandes13,XLuo15}.  Therefore, 
the Ising-nematic correlations, a state with no magnetic long-range order 
(staggered magnetization $M=0$) but with local broken $C_4$ symmetry lattice distortion \cite{CCL,si,jphu}, is the driving force for the observed
resistivity anisotropy \cite{fernandes11,CCL,si,jphu}. 

\begin{figure}[t]
\includegraphics[scale=.25]{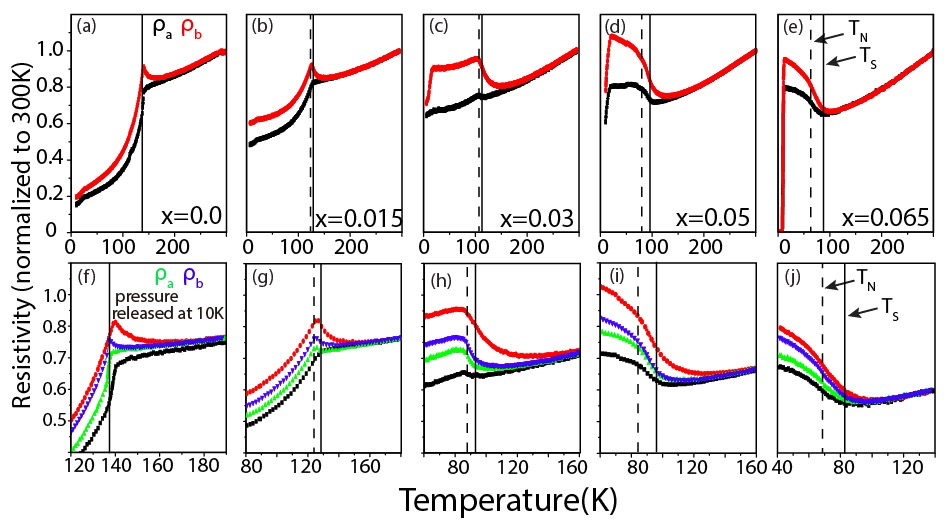}
\caption{ 
Temperature dependence of the resistivity anisotropy in strained and strain free BaFe$_{2-x}$Ni$_x$As$_2$.
(a-e) Temperature dependence of the in-plane resistivity $\rho_a$ (black) and $\rho_b$ (red) under uniaxial strain for 
BaFe$_{2-x}$Ni$_x$As$_2$ with $x = 0, 0.015, 0.03, 0.05, 0.065$, respectively. 
The vertical solid and dashed lines mark $T_s$ and $T_N$, respectively, for these materials without
uniaxial strain.
(f-j) Expanded view of the data in (a-e).  The green and blue data points are
$\rho_a$ and $\rho_b$ resistivity obtained on warming after 
releasing the pressure at 10 K. In all cases, the resistivity is measured on heating, with the same 
sample and same contacts (four point Montgomery method).
 }
 \end{figure}

We first compare transport measurements obtained on single domain samples detwinned using 
a standard mechanical clamp and the new device [Fig. 1(b)].  
The resistivity data along the orthorhombic $a$ and $b$ directions are measured 
via the Montgomery method \cite{Montgomery}.
Resistivity along the $a$ ($\rho_a$) and $b$ ($\rho_b$) directions
are measured in the same cycle using different current directions 
with wiring diagram shown in Fig. 1(d). Two sets of resistivity data as a function of temperature 
were collected for the detwinned crystals of 
BaFe$_{2-x}$Ni$_x$As$_2$.
Figure 2(a)-2(e) shows temperature dependence of $\rho_a$ and $\rho_b$ for 
$x=0, 0.015, 0.03, 0.05, 0.065$, respectively, under $\sim$10 MPa of uniaxial pressure.  Consistent with 
previous work \cite{fisher}, we see  
clear resistivity anisotropy ($\rho_b>\rho_a$) at temperatures above the strain free $T_N$ and 
$T_s$ marked as vertical dashed and solid lines, respectively.  
The green and blue lines in 
Figures 2(f)-2(j) show $\rho_a$ and $\rho_b$, respectively, on the warming cycle when 
the room-temperature applied strain is released at base temperature (10 K).  The corresponding 
$\rho_a$ and $\rho_b$ under uniaxial strain in Fig. 2(a)-2(e) are shown as black and red lines in 
Fig. 2(f)-2(j). 

\begin{figure}[t]
\includegraphics[scale=.50]{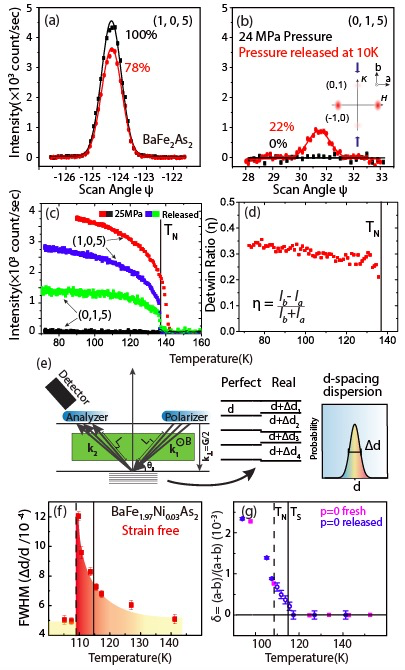}
\caption{Temperature dependence of Magnetic Bragg peaks at $(1,0,5)$ and $(0,1,5)$ 
in strained and strain free BaFe$_2$As$_2$ and neutron Larmor diffraction in BaFe$_{1.97}$Ni$_{0.03}$As$_2$.
(a) Transverse scan through magnetic Bragg peak $(1,0,5)$ with and 
without uniaxial strain at 10 K obtained
using flatcone setup and in-situ sample clamp on IN8.
(b) Identical scans through $(0,1,5)$ peak on IN8.
(c) Temperature dependence of the magnetic scattering at $(1,0,5)$ and $(0,1,5)$ in strained
and stress free case. (d) Estimated temperature dependence of the detwinning ratio $\eta$.
(e) Schematic diagram for the configuration of neutron 
Larmor diffraction measurements \cite{Rekveldt2001,Pfleiderer}.
In neutron Larmor diffraction, the neutron precession directions
are same in $L_1$ and $L_2$. It can accurately measure lattice spacing $d$ and its distortion $\Delta d$.
(f) Temperature dependence of the $\Delta d/d$.  The solid and dashed vertical lines are $T_s$ and $T_N$, respectively.
(g) Temperature dependence of the estimated orthorhombicity $\delta$ for fresh and strain free (first apply uniaxial pressure, 
then release pressure) BaFe$_{1.97}$Ni$_{0.03}$As$_2$. The most dramatic changes in lattice distortion happen at
$T_N$ and not at $T_s$. 
 }
\end{figure}

In the undoped parent compound ($x=0$), the uniaxial strain clearly increases the temperature below which 
the resistivity decreases with decreasing temperature [Fig. 2(f)], consistent with the notion that 
the uniaxial strain necessary for detwinning the sample also increases the $T_N$ of the system \cite{Dhital}.
In addition, we see that the uniaxial strain itself enhances the resistivity anisotropy both below and above 
$T_N$ ($T_s$).  Although much reduced, the resistivity anisotropy ($\rho_b>\rho_a$) 
is also present in the stress free tetragonal state above $T_N$ ($T_s$).  When the Ni-doping 
level is increased to $x=0.015, 0.03$, we find similar trend for strained and stress free 
resistivity [Figs. 2(g) and 2(h)].  Since $T_N$ and $T_s$ are now clearly separated, we can see that
the resistivity reduction in the stress free sample happens below $T_N$, and the resistivity anisotropy shows no
observable anomaly across $T_s$.  Upon further increasing the Ni-doping levels to 
$x=0.05, 0.065$, the resistivity smoothly increases on cooling across $T_s$ and no longer displays
a clear kink below $T_N$.  At all doping levels studied, we
find resistivity anisotropy in stress free samples above $T_N$ and $T_s$ (Fig. 2).
 
\begin{figure}[t]
\includegraphics[scale=.45]{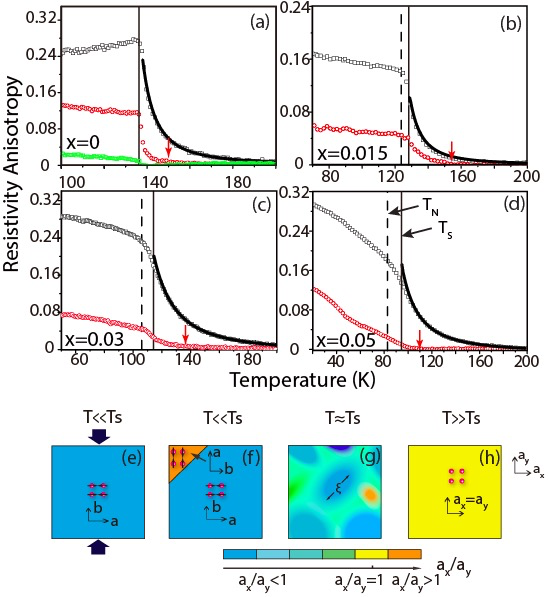}
\caption{
Temperature dependence of the in-plane resistivity anisotropy, defined as $2(\rho_b-\rho_a)/(\rho_b+\rho_a)$,  
for strained and strain free BaFe$_{2-x}$Ni$_{x}$As$_2$.
(a-d) Temperature dependence of the in-plane resistivity anisotropy 
for BaFe$_{2-x}$Ni$_{x}$As$_2$ under uniaxial pressure (Black) and pressure released at 10 K (Red) on warming. 
Solid and dashed vertical lines mark $T_s$ and $T_N$, respectively, for each Ni-doping. 
The solid black line is a fit above $T_s$ using Curie-Weiss functional form.  The red arrows mark the estimated 
$T^{\ast\ast}$.  The green curve in (a) represents resistivity anisotropy measured on cooling in stress free case
from room temperature.
(e-h) Microscopic picture of what happens in the process of releasing uniaxial strain at low temperature.
$a$ and $b$ are orthorhombic lattice parameters.  At temperatures slightly above $T_s$ in strain free 
case, the overall crystal structure is tetragonal but there are local orthorhombic lattice distortions induced by
the strong magnetoelastic coupling, which gives rise to the observed resistivity anisotropy. 
 }
\end{figure}

Although transport data in Fig. 2 revealed clear evidence for resistivity anisotropy in the stress free tetragonal state of 
underdoped BaFe$_{2-x}$Ni$_x$As$_2$ [Fig. 1(e)], these measurements cannot determine the sample detwinning ratio
upon releasing the uniaxial strain at low temperature 
 and microscopic origin
of the resistivity anisotropy above $T_s$. To address these questions, we carried out neutron diffraction experiments
on BaFe$_2$As$_2$ using an in-situ detwinning device similar to Fig. 1(b) and the 
flat-cone option of the IN8 triple-axis spectrometer at ILL \cite{kulda}.  In addition, we
performed neutron Larmor diffraction measurement on BaFe$_{1.97}$Ni$_{0.03}$As$_2$ using the TRISP triple-axis at Heinz Maier-Leibnitz,
Garching, Germany \cite{Rekveldt2001,Pfleiderer}.  We first describe neutron diffraction experiments on IN8 designed to study 
the detwinning ratio and its temperature dependence in strained and stress free BaFe$_2$As$_2$, as these results will allow us to 
determine if the detwinning ratio is maintained after releasing the uniaxial strain below $T_N$. 
For the experiment, an annealed square-shaped single crystal of BaFe$_2$As$_2$ ($\sim$220 mg) was mounted 
on a specially designed sample stick inside an orange cryostat. The momentum transfer ${\bf Q}$ in three-dimensional reciprocal space in \AA$^{-1}$ is defined as $\textbf{Q}=H\textbf{a}^\ast+K\textbf{b}^\ast+L\textbf{c}^\ast$, where $H$, $K$, and $L$ are Miller Indies and 
${\bf a}^\ast=\hat{{\bf a}}2\pi/a$, ${\bf b}^\ast=\hat{{\bf b}}2\pi/b$, ${\bf c}^\ast=\hat{{\bf c}}2\pi/c$ \cite{XYLu13a}.
In the AF ordered state of a 100\% detwinned sample, the AF Bragg peaks 
should occur at $(\pm 1,0,L)$ ($L=1,3,5,\cdots$) 
positions in reciprocal space and be absent at $(0,\pm 1,L)$. Our sample is aligned in the $[H,0,L]$ scattering plane.
Using the flatcone setup on IN8 \cite{kulda}, we can access both  $(1,0,5)$ and $(0,1,5)$ Bragg positions.
When a pressure of $\sim$24 MPa is applied along the $b$ direction of  BaFe$_2$As$_2$, the sample is 100\% detwinned 
with no magnetic scattering at (0, 1, 5) [Fig. 3(a) and 3(b)]. After releasing the uniaxial pressure at 10 K, we see that the sample becomes
partially twinned again with magnetic scattering intensity at both $I(1,0,5)$ and  
$I(0,1,5)$, giving a detwinning ratio of $\eta\approx56$\% ($\approx[I(1,0,5)-I(0,1,5)]/[I(1,0,5)+I(0,1,5)]$). 
This is consistent with transport measurements indicating a smaller resistivity anisotropy in the stress 
free BaFe$_2$As$_2$ [Fig. 2(f)]. 
Figure 3(c) shows temperature dependence of the magnetic scattering at 
$(1,0,5)$ and $(0,1,5)$ under 25 MPa uniaxial pressure and stress free.  
While the sample is 100\% detwinned under 25 MPa below $T_N$ with no magnetic scattering at   
$(0,1,5)$, the stress free sample has finite intensity at both
$(1,0,5)$ and $(0,1,5)$ below $T_N$.  Figure 3(d) shows temperature dependence 
of $\eta$, which reveals a decreasing detwinning ratio on warming to $T_N$.

In previous studies of the neutron extinction effect on   
the $(2,-2,0)$ nuclear Bragg peak of BaFe$_2$As$_2$ in zero pressure \cite{xylu14}, 
its intensity is found to deviate from normal behavior below $\sim$150 K before displaying a step like feature 
at $T_N\approx T_s\approx 138$~K, suggesting the presence of fluctuating orthorhombic structural domains above $T_s$.
Using neutron Larmor diffraction with polarized neutrons [Fig. 3(e)], we can precisely 
determine temperature dependence  
of the lattice parameter and its distortion \cite{Rekveldt2001,Pfleiderer}.
Since transport measurements in Fig. 2 suggest that the resistivity anisotropy in
stress free detwinned sample reduces dramatically above $T_N$ and shows no visible anomaly across
$T_s$ for lightly electron-doped BaFe$_{2-x}$Ni$_x$As$_2$ [Fig. 2(g) and 2(h)], we decided to study temperature dependence of the 
lattice distortions and orthorhombicity $\delta=(a-b)/(a+b)$ in BaFe$_{1.97}$Ni$_{0.03}$As$_2$ \cite{mgkim}, where 
$T_s$ and $T_N$ are well separated as determined from transport 
and neutron diffraction experiments.  For this purpose, we focus on (4,0,0) Bragg peak, which has a $d$-spacing $d=a/4$.
In a classical second order magnetic phase transition, one would
expect that spin-spin correlation length increases on cooling and diverges at 
$T_N$, while the underlying lattice correlations $\xi$ remain long-ranged and temperature independent.  
Surprisingly, our neutron Larmor diffraction measurements on stress free 
BaFe$_{1.97}$Ni$_{0.03}$As$_2$ reveal that the
lattice distortion ($\Delta d/d$) of the system shows no visible anomaly across $T_s$ ($\approx 118$ K), but increases 
continuously on cooling below $T_s$ before collapsing abruptly below $T_N$ ($\approx 109$ K) [Fig. 3(f)] \cite{supplementary}.
Similarly, instead of being a temperature independent constant, the lattice correlation length $\xi$ decreases on 
cooling, changing smoothly from 2500 \AA\ around $\sim$150 K to 1000 \AA\ just 
above $T_N$ with no anomaly across $T_s$ \cite{supplementary}.
Figure 3(g) compares temperature dependence of the lattice orthorhombicity $\delta$ for 
BaFe$_{1.97}$Ni$_{0.03}$As$_2$ without applying any external strain and in strain released sample.  In both cases, we see 
that AF order induces a large change in lattice orthorhombicity, consistent with previous X-ray scattering work \cite{mgkim}.
Therefore, BaFe$_{1.97}$Ni$_{0.03}$As$_2$ 
exhibits a strong magnetoelastic coupling near $T_N$.

Figure 4 summarizes temperature dependence of the resistivity anisotropy, 
defined as $2(\rho_b-\rho_a) /(\rho_b+\rho_a)$, for uniaxial strained and stress free
BaFe$_{2-x}$Ni$_x$As$_2$ with $x=0,0.015,0.03,0.05$. Similar to previous work \cite{chu12,HHKuo,HHKuo15}, we find that 
temperature dependence of the resistivity anisotropy 
in uniaxial strained samples can be well described by a Curie-Weiss functional form
above the strain free $T_s$ and below $T^\ast$ [see solid lines in Fig. 4(a)-4(d)].  When uniaxial strain is released, 
the resistivity anisotropy and its appearance temperature $T^{\ast\ast}$ are 
dramatically reduced. Nevertheless, it is clearly present above $T_s$ 
in the tetragonal phase.  For strain free samples cooled from high-temperature paramagnetic tetragonal phase,
there are no resistivity anisotropy above $T_s$ [see green data points in Fig. 4(a)].  The small resistivity anisotropy below $T_N$ is due to slight imbalance in the twin domain populations. 

To understand the observed resistivity anisotropy behavior, we consider a 
microscopic scenario as shown in Fig. 4(e)-4(h).  In the low-temperature uniaxial strained detwinned state, 
the undoped and underdoped BaFe$_{2-x}$Ni$_x$As$_2$ form a single domain homogeneous magnetic ordered state with 
intrinsic resistivity anisotropy that is weakly electron-doping dependent [Fig. 4(a)-4(d), $\rho_b>\rho_a$]. 
Upon releasing the uniaxial strain, the sample becomes partially detwinned AF ordered state
 with reduced resistivity anisotropy [Fig. 4(f)].
 On further warming to temperatures above $T_N$ and $T_s$, 
these materials exhibit a large 
lattice distortion across $T_N$ but much less anomaly across $T_s$ [Fig. 3(f) and 3(g)] \cite{mgkim}.
These results suggest that the resistivity anisotropy seen in the narrow temperature region above $T_s$
is due to the remnant local lattice distortions arising from the 
large magnetoelastic coupling across $T_N$ [Fig. 4(g)].  The system finally relaxes to the true homogeneous tetragonal
state without resistivity anisotropy at temperatures above $T^{\ast\ast}$.  Since 
our neutron Larmor diffraction measurements showed 
no additional anomaly in lattice parameters and lattice distortion above $T_s$, we conclude that 
there is no thermodynamic phase transition at $T^\ast$ and $T^{\ast\ast}$ 
in agreement with recent heat capacity measurements \cite{XLuo15}.  The resistivity anisotropy seen in the stress free
detwinned samples below $T^{\ast\ast}$ on warming across $T_N$ is then 
due to local spin nematic correlations and associated lattice distortions arising from the magnetoelastic 
coupling through the collinear AF state below $T_N$. The absence of such effect in strain free sample 
on cooling confirms this scenario and the weakly first order nature of the magnetic transition.

In summary, by using a specially designed in-situ detwinning device,
we have discovered the presence of resistivity anisotropy in the tetragonal phase of stress free
BaFe$_{2-x}$Ni$_x$As$_2$ below $T^{\ast\ast}$, a temperature lower than $T^\ast$ 
associated with resistivity anisotropy in uniaxial strained sample \cite{jhchu,matanatar,fisher,chu12}. 
Our neutron diffraction experiments confirm the partially detwinned state in the stress free sample, thus 
indicating that the observed resistivity anisotropy arises from local
spin nematic correlations and lattice distortions.  Furthermore, our neutron Larmor
diffraction experiments on lightly electron-doped BaFe$_{1.97}$Ni$_{0.03}$As$_2$ indicate 
lattice distortions across $T_N$ and $T_s$ with no evidence of another phase transition above $T_s$.  
These results thus establish that resistivity anisotropy in the tetragonal phase arises 
from the magnetoelastic coupling associated with static AF order, suggesting the presence of  
local Ising-nematic spin correlations and lattice distortions 
in the tetragonal state of electron-doped iron pnictides near $T_N$.

We are grateful to Sebastien Turc, E. Bourgeat-Lami, E. Leli$\rm \grave{e}$vre-Berna of ILL, France for designing and 
constructing the detwinning device used at IN8. 
The transport and neutron work at Rice is supported by the
U.S. NSF-DMR-1362219 and DMR-143606 (P.D.).  This work is also supported by 
the Robert A. Welch Foundation Grant Nos. C-1839 (P.D.) and C-1411 (Q.S.).
Q.S. is supported by the U.S. NSF-
DMR-1309531.

\newpage
\appendix

\section{Supplementary Materials for: Electronic nematic correlations in the stress free tetragonal state of BaFe$_{2-x}$Ni$_{x}$As$_{2}$}

\subsection{Sample Information}

The BaFe$_{2-x}$Ni$_{1-x}$As$_{2}$  single crystals were grown using self-flux method as described before \cite{ycchen}.
The crystal orientation was determined by X-ray Laue machine, and the square shaped samples were cut for Montgomery method resistivity measurements. The samples were annealed at 800 K for 2 days to reduce defects and disorder.  

\subsection{Resistivity measurement: Montgomery method}

For sheet like samples, measurement of anisotropic in-plane resistivity can be 
carried out by Montgomery method \cite{Montgomery:2003bn}. 
The samples are cut along the $a$ and $b$ axes directions 
into a squared shape with the $c$-axis perpendicular to the squared surface. 
Current is applied through contacts at two adjacent corners of the planar face and the potential is measured at the other two corners at the same plane [SFig 1.(a)]. From the measurements, we can get $R_1=V_1/I_1$. Similar measurements can be done with the electrical connections rotated 90$^\circ$ with respect to the original setup [SFig1.(b)], which gives $R_2=V_2/I_2$. 
This will allow calculations of the resistivity anisotropy.

Results from an anisotropic sample with dimensions $l_1$, $l_2$, $l_3$, and resistivity $\rho_1, \rho_2, \rho_3$ can be estimated from an isotropic sample with dimension  $l^\prime_1$, $l^\prime_2$, $l^\prime_3$ with the transformation:
\begin{equation}
\rho^3=\rho_1 \rho_2 \rho_3,
\end{equation}
and
\begin{equation}
l^\prime_i=l_i(\rho_i/\rho)^{1/2}. 
\end{equation}

Through theoretical calculations, we can get for small $l_1$/$l_2$ ratio:
\begin{equation}
\frac{l^\prime_2}{l^\prime_1}\simeq \frac{1}{2}\left[ \frac{1}{\pi}ln\frac{R_2}{R_1}+\sqrt{\left[ \frac{1}{\pi}ln\frac{R_2}{R_1}\right] ^2+4} \right]
\end{equation}
and the anisotropic resistivity in the plane can be written as:
\begin{equation}
\rho_1=A (l^\prime_1/l^\prime_2)R_1 \sinh[\pi l^\prime_2/l^\prime_1]
\end{equation}
and 
\begin{equation}
\rho_2=B (l^\prime_2/l^\prime_1)R_2 \sinh[\pi l^\prime_1/l^\prime_2].
\end{equation}

In the case of BaFe$_{2-x}$Ni$_{x}$As$_{2}$ , $A$ and $B$ are pre-factors 
that can be normalized as $\rho_1(300\ K)=\rho_2(300\ K)$. Then $\rho_1$ 
and $\rho_2$ can be determined by resistance measurement $R_1$ 
and $R_2$ by two different channel on PPMS resistivity puck, as shown in SFig 1.(c) and (d). $\rho_1$ 
and $\rho_2$ derived from Montgomery method is the equivalent resistivity by the mixture of two domains angling $90^o$ with initial resistivity of $\rho_{a0}$ 
and $\rho_{b0}$.

\begin{figure}
\includegraphics[width=8cm]{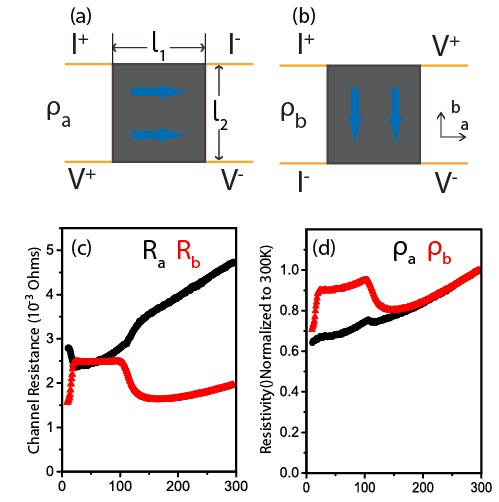}
\caption{(a),(b) Schematic illustration of Montgomery method: wire connections and current flow directions.
 Sample plane dimensions are $l_1 \times l_2$ .
 The blue arrows on the sample are the current direction for each set up. 
(c) Resistance data of BaFe$_{1.97}$Ni$_{0.03}$As$_{2}$ from the two channels on the PPMS resistivity puck. 
Black and Red indicate the direction of the resistance measured on the channel. (d) Normalized resistivity calculated from (c). }
\label{fig:1}
\end{figure}

\subsection{Pressure Dependence Measurements of the resistivity anisotropy}

The pressure dependence of the in-plane resistivity was studied systematically both as a function of Ni-doping and temperature. 
The magnitude of the uniaxial pressure applied to the system is determined approximately by the 
length compression of the spring as shown in Fig. 1(b) of the main text.
In SFig. 2(a), we show pressure dependence of the resistivity
anisotropy defined as $2(\rho_b-\rho_a)/(\rho_a+\rho_b)$ for zero pressure cooled and pressure cooled case of BaFe$_2$As$_2$.  The pressure dependence of the
detwinning ratio is also plotted.  In the zero pressure cooled case, the sample was cooled down to 10 K with no pressure applied and then temperature was raised to the targeted temperature.
In the pressure cooled case, the sample was cooled with maximum pressure (~15 MPa) to 10 K, then the pressured is released at 10 K and temperature was raised to the targeted temperature.
 To the first order approximation, the resistivity anisotropy 
tracks the detwinning ratio of the system before the sample is fully detwinned.  
SFigure 2(b) shows similar data at 137 K.
SFigure 2(c) shows pressure dependence of the resistivity anisotropy 
across the AF ordering ($T_N$) and structural ($T_s$) transitions.
For temperatures above $T_N$ and $T_s$, the pressure and resistivity anisotropy 
relationship becomes linear and the slope decreases with increasing 
temperature, consistent with previous work \cite{HHKuo15}. In SFigure 2(d), 
we plot the pressure dependence of the resistivity anisotropy at 
different temperatures below and above $T_N$ and $T_s$ for 
 BaFe$_{1.985}$Ni$_{0.015}$As$_2$.  Clear hysteresis is seen in the data, suggesting a partially detwinned sample after 
releasing the pressure.
For the pressure released partially detwined sample, the resistivity anisotropy and the detwinning ratio follow the same trend before $T_N$,
as shown in SFig. 2(a),  which suggests the proximate proportionality of resistivity anisotropy with 
the detwinning ratio $\eta$.  SFigure 3(a) compares temperature dependence of the detwinning ratio $\eta$ with that of the resistivity
anisotropy.  The remarkable similarity in these data again confirms the notion that the reduced
resistivity anisotropy in stress free sample is due to reduced detwinning ratio.

\begin{figure}
\includegraphics[width=8cm]{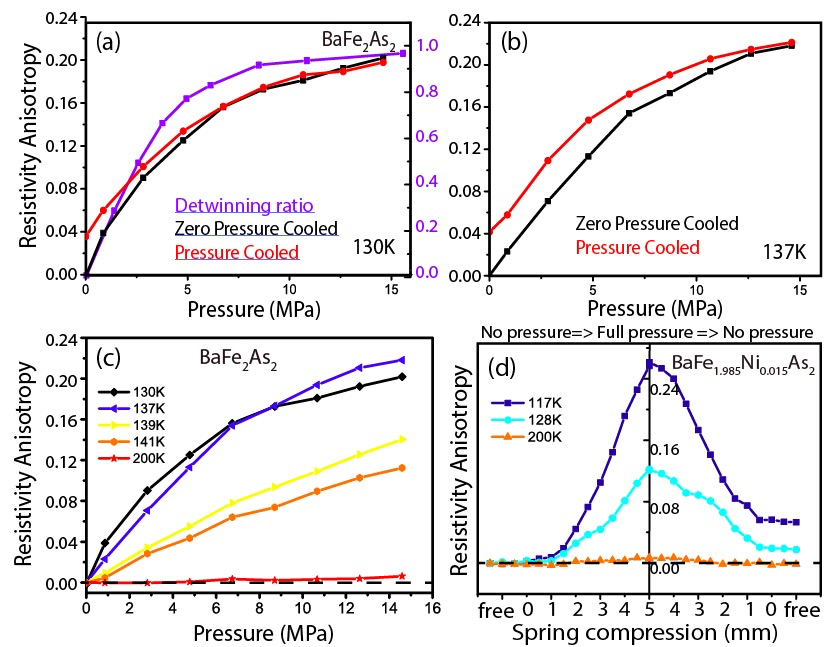}
\caption{ (a),(b) Resistivity anisotropy of BaFe$_2$As$_2$ for the zero pressure cooled (Black) and pressure cooled (Red) sample. Purple line in (a) is the detwinning ratio of BaFe$_2$As$_2$ determined by elastic neutron scattering using IN8.
(c) pressure dependance of resistivity anisotropy in BaFe$_2$As$_2$ (zero pressure cooled) at different temperature.
(d)Detwinning effect in BaFe$_{1.985}$Ni$_{0.015}$As$_2$ for zero pressure cooled sample at different temperatures.
}
\label{fig:2}
\end{figure}

\subsection{Neutron Larmor Diffraction Experiments}

Neutron Larmor diffraction is an ideal technique for measuring lattice distortion and expansion, with a resolution better than $10^{-5}$ for $\Delta d/d$. The resolution is not affected by sample mosaicity or slight sample misalignment, enabling us to keep track of small changes in lattice spacing $d$ and its distortion $\Delta d$. Detailed principles of Larmor diffraction technique can be found in references \cite{ldepl,keller2,LDscience}. 

To determine the lattice orthorhombicity induced Bragg 
peak splitting in the unstrain sample between $T_N$ and $T_s$, we assume the 
full-width-half-maximum (FWHM) of the peaks is unchanged across $T_s$. The result is described in Fig. 3 of the main text.

As discussed in the main text, for a standard second order magnetic phase transition, one would 
expect that lattice distortion of the system ($\Delta d/d$) remains unchanged across $T_N$.  Since this is clearly not the case
for electron underdoped BaFe$_{1.97}$Ni$_{0.03}$As$_2$, one can estimate temperature dependence of 
the lattice correlation length $\xi$, defined as Fourier transform of the Bragg peak width determined using Larmor diffraction \cite{xylu13}. 
Assuming that the $d$-spacing spread follows a Gaussian distribution, the FWHM of of its Fourier transform gives 
the lattice correlation length.  For typical triple-axis experiment, the instrument resolution is about 300 \AA.  Here the resolution is
much better as shown in temperature dependence of the lattice correlation length $\xi$ in SFig. 3(b).
As a function of decreasing temperature, the lattice correlation length reduces from 
2500 \AA\ at 150 K to 1000 \AA\  at 110 K before becoming 2500 \AA\  again in the AF ordered state.

\begin{figure}
\includegraphics[width=7.0cm]{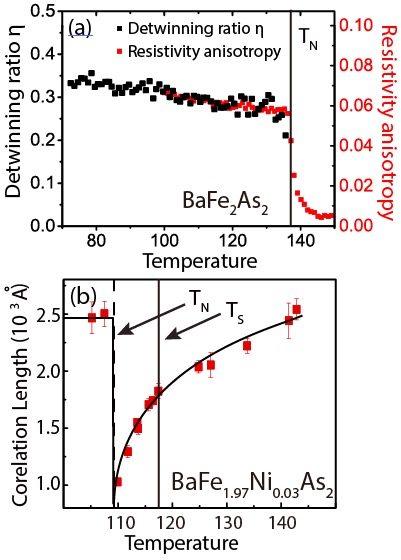}
\caption{ (a)Temperature dependance of detwinning ratio $\eta$ (black) and resistivity anisotropy in stress free BaFe$_2$As$_2$. 
(b)Temperature dependance of the lattice correlation length $\xi$ in BaFe$_{1.97}$Ni$_{0.03}$As$_2$ as determined from neutron 
Larmor diffraction.
}
\label{fig:3}
\end{figure}

\end{document}